\documentclass[11pt]{article}

\usepackage[utf8]{inputenc}
\usepackage[T1]{fontenc}
\usepackage[english]{babel}

\usepackage{indentfirst}
\usepackage{amsmath}
\usepackage{amssymb}
\usepackage{graphicx}
\usepackage{subfigure}
\usepackage{psfrag}
\usepackage{color}

\allowhyphens

\newcommand{\bml}{\begin{mathletters}}
\newcommand{\eml}{\end{mathletters}}
\newcommand{\be}{\begin{equation}}
\newcommand{\ee}{\end{equation}}
\newcommand{\ba}{\begin{array}}
\newcommand{\ea}{\end{array}}
\newcommand{\bea}{\begin{eqnarray}}
\newcommand{\eea}{\end{eqnarray}}
\newcommand{\nn}{\nonumber}

\newcommand{\pr}{\prime}
\newcommand{\dpr}{\prime\prime}

\newcommand{\valos}{\mathbb{R}}

\newcommand{\eps}{\varepsilon}

\newcommand{\ordo}{\mathcal{O}}

\newcommand{\ket}[1]{{\left|#1\right\rangle}}
\newcommand{\bra}[1]{{\left\langle #1\right|}}

\newcommand{\skalarszorzat}[2]{{\langle #1 | #2 \rangle}}

\usepackage{ifpdf}

\ifpdf
\usepackage{epstopdf}
\usepackage[pdftex,ps2pdf,dvips,colorlinks,urlcolor=blue,citecolor=blue,linkcolor=blue]{hyperref}
\else
\usepackage[hypertex,ps2pdf,dvips,colorlinks,urlcolor=blue,citecolor=blue,linkcolor=blue]{hyperref}
\fi

\makeatother

\newcommand{\MJ}{\mathcal{J}}
\newcommand{\MD}{\mathcal{N}}
\newcommand{\MB}{\mathcal{B}}

\newcommand{\MN}{\mathcal{A}}

\begin{document}

~ \vspace{0.5cm}

\begin{center} \vskip 14mm
{\Large\bf On $\mathcal{O}(1)$ contributions to the free energy in
  Bethe Ansatz systems: the exact $g$-function}\\[10mm] 
{\large 
Bal\'azs Pozsgay~~\footnote{Email: 
{\tt pozsgay.balazs@gmail.com}}}
\\[8mm]
Institute for Theoretical Physics, Universiteit van Amsterdam,\\
 Valckenierstraat 65, 1018 XE Amsterdam, The Netherlands
\vskip 22mm
\end{center}

\begin{quote}{\bf Abstract}\\[1mm]
We investigate the
sub-leading contributions to the free
energy of Bethe Ansatz solvable (continuum) models with different boundary conditions. 
We show that the Thermodynamic Bethe Ansatz approach
is capable of providing the $\ordo(1)$ pieces
if both the density of states in rapidity space and the quadratic fluctuations around the
saddle point solution to the TBA are properly taken into account. 
In relativistic boundary QFT the $\ordo(1)$
contributions are directly related to the exact $g$-function.
In this paper we provide an all-orders proof
of the previous results of P. Dorey et al. on the $g$-function in
both massive and massless models.
In addition, we derive a new result for the $g$-function which
applies to massless theories with arbitrary diagonal scattering in
the bulk.
\end{quote}

\vfill

\numberwithin{equation}{section}

\newpage

\tableofcontents

\section{Introduction}

The study of the thermodynamics of one dimensional integrable models
with factorized scattering dates back to the seminal work of Yang and
Yang \cite{YangYang1,YangYang2}. Their method, known today as the Thermodynamic
Bethe Ansatz (TBA) is quite general and it was worked out for a
large number of models relevant to condensed matter physics
\cite{Takahashi-book}. In its simplest formulation the TBA is written
down for periodic boundary conditions and it provides the free energy
density, ie. the $\ordo(L)$ part of the free energy. 

In this paper we
study the sub-leading  pieces of the free energy for different boundary
conditions in continuum models. We restrict ourselves to theories with
diagonal scattering; however it is
  expected, that the proposed methods will work even in the non-diagonal
case \footnote{see the related comments in the Conclusions
  (Sec. \ref{concl})}. 
In this paper we focus mainly on relativistic theories. 

In integrable relativistic field theory the TBA was introduced by
Al. Zamolodchikov 
in \cite{zam-tba}; soon thereafter it became one of the central tools to
study finite size effects.  In relativistic models Euclidean
invariance implies that the free energy density at finite 
temperature 
is directly related to the
exact ground state energy in finite volume. 
Studying this quantity it is possible to recover the behaviour around the fixed points  of the
renormalization group flow, which are usually given by a conformal field
theory. 
Thus the TBA provides a link between the scattering theory
(IR) and perturbed CFT (UV) description of the same model:
it predicts the central charge, the scaling dimensions of the perturbing operator,
and various other quantities
\cite{zam-tba,klassen_melzer_tba1,klassen_melzer_tba2}. 

The techniques of integrability can be applied to problems
with non-trivial integrable boundary conditions; in relativistic
scattering theory the foundations were laid down in
\cite{Ghoshal:1993tm}. One object of particular interest is the exact
$g$-function, which is the off-critical generalization of the
non-integer ground state degeneracy of critical boundary conditions
 introduced by Affleck and Ludwig in the context of the Kondo
model \cite{Affleck:1991tk}. The 
$g$-function describes the $\ordo(1)$ contribution of a single boundary to
the free energy 
and
it can be used to study renormalization group flows in
the space of boundary field theories
\cite{Dorey:1997yg,Dorey:1999cj}. 
In
\cite{Friedan:2003yc} it was shown that the $g$-function satisfies a
gradient formula, from which it follows that in unitary theories the
boundary entropy 
monotonically decreases under the RG flow. This is the
$g$-theorem, which can be regarded as the boundary-counterpart of the
celebrated $c$-theorem by A. B. Zamolodchikov \cite{Zamolodchikov:1986gt}. 

It is an old and very  natural idea 
to determine the $g$-function in the framework
of the Thermodynamic Bethe Ansatz. 
The first result appeared in
\cite{LeClair:1995uf}, where the authors proposed a simple formula
based on the boundary-dependence of the Bethe equations.
Later it was found in
 \cite{Dorey:1999cj}  that although the
results of  \cite{LeClair:1995uf} correctly describe the
boundary-dependence of the $g$-function, a
boundary-independent term has to be added in order to match the predictions of CFT. The missing
piece was derived in \cite{Dorey:2004xk} using a cluster expansion for
the free energy; the exact result was expressed in terms
of the solution of the TBA with periodic boundary conditions. While this
exact $g$-function successfully passed a number of non-trivial tests
\cite{Dorey:2004xk,Dorey:2005ak} and recently it was generalized to
describe a
massless flow in \cite{Dorey:2009vg}, the interpretation of the
boundary-independent terms remained unclear. 

A remarkable attempt to obtain the non-extensive pieces to the free energy
of Bethe Ansatz systems
was performed in \cite{woynarovich}, where it was shown that the
quadratic fluctuations around the saddle point solution yield a
well-defined $\ordo(1)$ piece. However, the calculation of
\cite{woynarovich} seemed to contradict all previous results: it did
not reproduce the boundary-independent term of \cite{Dorey:2004xk}, moreover it
predicted an $\ordo(1)$ piece even in the periodic case where no such
term is expected.

In this paper we revisit the calculations of \cite{woynarovich} and
argue that the only
flaw of \cite{woynarovich} is
that it did not take into account the non-trivial density of states in
the configuration space of Bethe Ansatz systems. In other words, 
the functional integral for the partition function was built on an incorrect
integration measure. 
We propose a 
 new normalization based on the thermodynamic behaviour of the density
 of states and we obtain the correct results in all
previously considered cases.

The paper is organized as follows. In the next subsection we provide the
necessary definitions for the $g$-function of relativistic boundary
field theory. Sec. 2 serves as a warm-up: we consider general Bethe Ansatz systems and
the behaviour of the density of states in the
thermodynamic limit.
In. 2.2 we revisit the calculations of \cite{LeClair:1995uf} and show
that it is possible to obtain the boundary-independent part of the
$g$-function 
by a simple heuristic argument. Motivated by these
findings in Sec. 3 we present
a general framework to evaluate all $\ordo(1)$ pieces to the free
energy. These formal results are then evaluated explicitly in massive and
massless relativistic models in Sections 4 and 5,
respectively. Finally Sec. 6
includes our conclusions.

\subsection{The exact $g$-function -- definitions}

The exact $g$-function can be defined as follows
\cite{LeClair:1995uf,Dorey:2004xk}.
 We restrict ourselves to
the simplest case with only one massive particle in the spectrum; the
generalization to other models is straightforward.

Let us consider a finite cylinder with height $L$ and circumference
$R$ (fig. 1). The integrable boundaries $a$ and $b$ are placed on the
two ends of the cylinder. 
\begin{figure}
  \centering
\psfrag{L}{$L$}
\psfrag{R}{$R$}
\psfrag{Hp}{$H_{ab}(L)$}
\psfrag{Hb}{$H(R)$}
\psfrag{Ba}{$\ket{B_a}$}
\psfrag{Bb}{$\ket{B_b}$}
\subfigure[]{\includegraphics[scale=0.5]{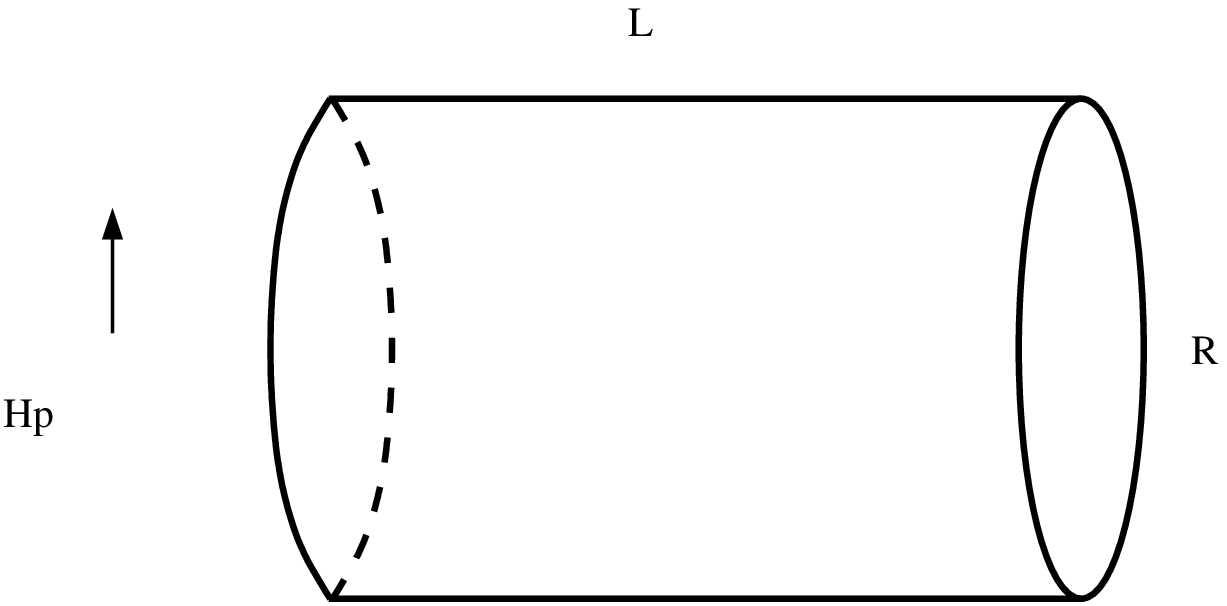}\label{bu}}
\subfigure[]{\includegraphics[scale=0.5]{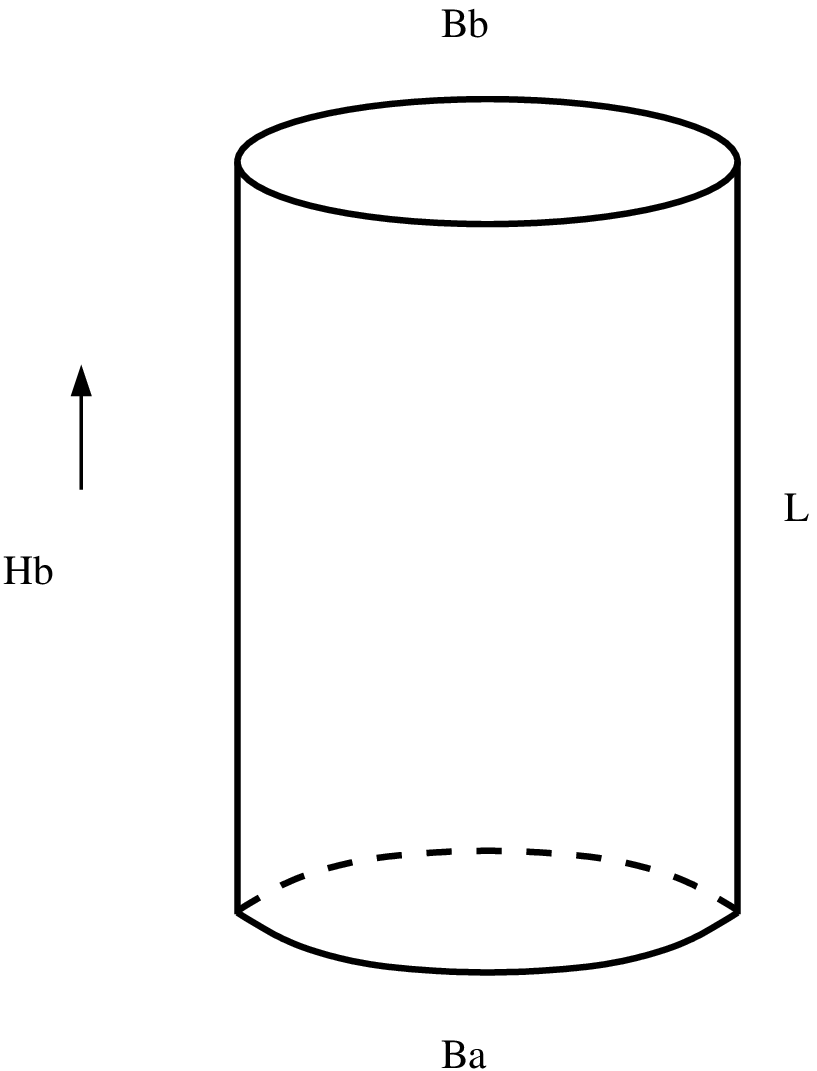}\label{hu}}
\caption{Pictorial representation of the two channels for the
  evaluation of the cylinder partition function.
The (imaginary) time evolution is generated in the vertical direction
by the corresponding Hamiltonians.}
\end{figure}
The partition function can be evaluated in two different channels.
Viewing $R$ as the direction of time (fig. \ref{bu}) one obtains
\begin{equation}
\label{Z1}
  Z_{ab}(R,L)=\text{Tr}\  e^{-H_{ab}(L)R}=\sum_\psi e^{-E_{\psi}^{ab}(L) R},
\end{equation}
where $H_{ab}(L)$ is the Hamiltonian of the open system of size $L$
with integrable boundary conditions described by the reflection
factors $R_{a}(\theta)$ and $R_{b}(\theta)$. The summation runs over a complete set of
states and $E_{\psi}^{ab}(L)$ are the eigenvalues of the Hamiltonian $H_{ab}(L)$.

On the other hand, one can use $L$ as the
time variable (fig. \ref{hu}). In this picture the boundaries play the role of initial
and final states of the time-evolution operator $H(R)$, which is the
Hamiltonian of the system of size $R$ with periodic boundary
conditions. In this channel the partition function is evaluated as
\begin{equation}
\label{Z2}
  Z_{ab}(R,L)=\bra{B_a}  e^{-H(R)L}\ket{B_b}=\sum_\psi \big(G_a^\psi(R)\big)^*G_b^\psi(R)e^{-E_\psi(R) L}
\end{equation}
Here $\ket{B_{a}}$ and $\ket{B_{b}}$ are the boundary states
corresponding to the boundary conditions \cite{Ghoshal:1993tm} and the
summation runs over a complete set of states of the periodic
bc. system. The amplitudes $G_{a,b}^\Psi(R)$ are defined as the
normalized overlaps
\begin{equation}
\label{djvu}
  G_j^\psi(R)=\frac{\skalarszorzat{\psi}{B_j}}
{\sqrt{\skalarszorzat{\psi}{\psi}}},\qquad j=a,b
\end{equation}
Equations \eqref{Z1} and \eqref{Z2} have to be contrasted with the
definition of the partition function with periodic boundary conditions
in both direction:
\begin{equation}
  \label{ZP}
  Z(R,L)=\text{Tr}\  e^{-H(R)L}
=\sum_\psi e^{-E^\psi(R) L}=Z(L,R)
\end{equation}

If $mL\gg 1$ then \eqref{Z2} and \eqref{ZP} are dominated by the ground state with
energy
\begin{equation*}
  E_0(R)=\epsilon R+\ordo(e^{-mR}),
\end{equation*}
where $\epsilon$ is the bulk energy density, which can be determined
by comparing the TBA results to conformal perturbation theory
\cite{zam-tba,klassen_melzer_tba1,klassen_melzer_tba2}.

Similarly, if  $mR\gg 1$ then \eqref{Z1} is dominated by the ground
state of the boundary system with energy
\begin{equation*}
  E_0^{ab}(L)=\epsilon L+f_a+f_b+\ordo(e^{-mL}),
\end{equation*}
where $f_a$ and $f_b$ are non-extensive boundary-contributions which can be
obtained from the boundary TBA \cite{LeClair:1995uf,Dorey:1997yg}.

Comparing \eqref{Z1} and \eqref{Z2} in the regime $mL,mR\gg 1$ one
finds
\begin{equation*}
  G_j^0(R)=e^{-f_jR}\Big(1+\ordo(e^{-mR})\Big),\qquad j=a,b
\end{equation*}
The $g$-function is then traditionally defined as
\begin{equation*}
  G_j^0(R)=e^{-f_jR}g_j(R),\qquad j=a,b
\end{equation*}
It follows from dimensional arguments that the $g$-function depends on
$r=mR$ only. It is useful to re-define the 
partition functions as
\begin{equation}
\label{ZP2}
\tilde Z(R,L)=  \sum_\psi e^{-(E_\psi(L)-\epsilon L) R}
\end{equation}
\begin{equation}
\label{Z3}
\tilde Z_{ab}(R,L)=  \sum_\psi e^{-(E_\psi^{ab}(L)-f_a-f_b-\epsilon L) R}
\end{equation}
With this prescription the vacuum energies have the asymptotics
\begin{equation*}
  \lim_{L\to\infty} E_0^{ab}(L)= \lim_{L\to\infty} E_0(L)=0,
\end{equation*}
and the excited state energies are calculated additively in the
Bethe Ansatz picture.
The $g$-function is then given by the limit
\begin{equation}
\label{g-fuggveny}
  \log g_a(r)g_b(r)=\lim_{L\to\infty} \left(\log \tilde Z_{ab}(R,L)
- \log \tilde Z(R,L)
\right)
\end{equation}
Our goal is the evaluation of the partition functions \eqref{ZP2} and \eqref{Z3} in the framework of
the Thermodynamic Bethe Ansatz, with a special emphasize on the
overall normalization. The exact $g$-function will be determined by the
relation above. 

We would like to remind the reader that in relativistic scattering
theories the Bethe Ansatz does not provide an exact description of the
spectrum. In fact there are exponentially decaying
residual finite size effects, which modify the multi-particle energies
obtained from the Bethe Ansatz \cite{klassen_melzer,Bajnok:2008bm}. In the conventional derivation of the
TBA these contributions are neglected; this is certainly a good
approximation in the dilute regime, where the average distance between
the particles is 
much larger than their Compton-wavelengths. However, it can be argued
that the TBA yields the correct results at any
temperatures and any densities \cite{niedermayer-On}.
In this paper we
develop arguments about the thermodynamic behaviour of certain Bethe
Ansatz quantities. We do not use any special assumptions other than
those already used in the derivation of the TBA. Therefore we may
neglect exponential corrections; this assumption will be justified by
the results, which are in agreement with the CFT results in the UV
(high temperature) limit.

\section{Thermodynamic Bethe Ansatz and density of states}

\label{difsetup}

Let us consider Bethe Ansatz systems with one particle type and no
internal degrees of freedom. We use the rapidity variable
$\theta$ to parametrize the states. The scattering is assumed to be
elastic and factorizing; the two-particle scattering is described by
the pure phase
\begin{equation*}
S(\theta_i-\theta_j)=e^{i\vartheta(\theta_i-\theta_j)}
\end{equation*}
Energy and momentum are given by
the functions $e(\theta)$ and $p(\theta)$. In relativistic field
theories
\begin{equation*}
  e(\theta)=m\cosh\theta \qquad p(\theta)=m\sinh\theta
\end{equation*}
whereas in the non-relativistic case we have
\begin{equation*}
  e(\theta)=\frac{m}{2}\theta^2 \qquad p(\theta)=m\theta
\end{equation*}
Alternatively, one can introduce a chemical potential according to
\begin{equation*}
  e(\theta)\quad\rightarrow \quad e(\theta)-\mu
\end{equation*}
In the relativistic case we always set $\mu=0$.

Consider the Bethe-Yang quantization of an $N$-particle state in
finite volume $L$ with periodic boundary conditions:
\begin{equation*}
  e^{ip_jL}\prod_{k\ne j}S(\theta_j-\theta_k)=1,\qquad
  j=1\dots N,
\end{equation*}
where $p_j=p(\theta_j)$. In the logarithmic form:
\begin{equation}
\label{logBY}
  Q_j=p_jL+\sum_{k\ne j}\vartheta(\theta_j-\theta_k)=2\pi
  I_j\qquad j=1\dots N,\quad I_j\in \mathbb{Z}
\end{equation}
We assume that the quantum numbers $\{I_1,\dots,I_N\}$ completely
characterise the state and that the
solutions consist of purely real rapidities, ie. we do not consider theories
with string-like solutions. Moreover, we assume that the solutions of
\eqref{logBY} span the whole Hilbert-space. 

The multi-particle energies can be calculated additively:
\begin{equation*}
  E\big[\{I_1,\dots,I_N\}\big]=\sum_{j=1}^N e(\theta_j)
\end{equation*}

The solutions of \eqref{logBY} are placed evenly in the space of the
quantum numbers. In rapidity space one can  define the density of
states  as
\begin{equation*}
  \rho_N(\theta_1,\dots,\theta_N)=\det \mathcal{J},\qquad
  \mathcal{J}_{ik}=\frac{\partial Q_i}{\partial \theta_k}
\end{equation*}
Alternatively, this can be interpreted as the norm of the Bethe Ansatz
state \cite{korepinBook}.

In the presence of integrable boundaries the quantization conditions can be
written as
\begin{equation}
  \label{BA-R0}
  e^{i2p_jL} R_a(\theta_j)R_b(\theta_j) \prod_{k\ne j}
  S(\theta_j-\theta_k) S(\theta_j+\theta_k)=1,
\qquad j=1\dots N,\quad\theta_j>0
\end{equation}
or
\begin{equation*}
\bar Q_j=  2p_jL+\vartheta_a(\theta_j)+\vartheta_b(\theta_j)+
\sum_{k\ne j}  \vartheta(\theta_j-\theta_k)+\vartheta(\theta_j+\theta_k)=
2\pi I_j, \qquad j=1\dots N,\quad\theta_j>0
\end{equation*}
where $R_{a,b}(\theta)=e^{i\vartheta_{a,b}(\theta)}$ are the elastic reflection
factors of the boundaries. The density of states is defined as
\begin{equation*}
\bar\rho_N(\theta_1,\dots,\theta_N)=\det \bar{\mathcal{J}},\qquad
 \bar{\mathop{\mathcal{J}}}_{ik}=\frac{\partial \bar Q_i}{\partial \theta_k}
\end{equation*}
We will be interested in the thermodynamic behaviour of 
$\rho_N$ and $\bar\rho_N$. These quantities will play an important
role in the normalization of the partition function in 
\ref{RolfMaierBode} and in the general treatment of section \ref{fasza}.

If the particle number is kept fixed and the volume is sent to
infinity, then the densities behave as $L^N$. In particular
\begin{equation*}
\rho_N,\bar\rho_N\quad 
\sim\quad  \Big(1+\ordo(L^{-1})\Big) \times \prod_{j=1}^N \big(2\pi \sigma(\theta_j)L\big)
\end{equation*}
where $\sigma(\theta)=p'(\theta)/2\pi$. Note that although the leading
piece is the same for $\rho_N$ and $\bar\rho_N$, the sub-leading terms
are in general different. An important property is that the coefficient of
these  $\ordo(L^{-1})$ term grows linearly with
$N$. Therefore it is expected that if  both $L$ and
$N$ go to infinity with their ratios fixed, then the densities will have a
non-trivial thermodynamic limit, which is expected to be different for
$\rho_N$ and $\bar\rho_N$. In this limit all higher order terms of
$\ordo(L^{-n})$ contribute.

In the following we consider the particular
case when the state in question is a finite-volume micro-canonical
realization of an infinite volume thermal state. In other words, we
send $N$ and $L$ to infinity with their ratios fixed, and we choose
the distribution of Bethe-Ansatz roots according to some Thermodynamic
Bethe Ansatz equation. 

As a starting point we briefly review the basic
notions of the TBA.
In the thermodynamic limit
we introduce the total density of
states\footnote{We use the term ``density'' for
  both $\rho_N$ and $\rho(\theta)$, $\rho^{(o)}(\theta)$. However, these quantities
  have a very different meaning: $\rho(\theta)$ and $\rho_o(\theta)$
  describe the distribution of roots {\it within one configuration},
  whereas $\rho_N$ describes the density of states {\it in the
  (thermodynamic dimensional) configuration space.}} $\rho(\theta)$ and also the density of occupied states
$\rho^{(o)}(\theta)$. The particle
density is given by the integral
\begin{equation*}
  n=\frac{N}{L}=\int d\theta\ \rho^{(o)}(\theta)
\end{equation*}
In the periodic case the densities satisfy the constraint
\begin{equation}
\label{Lieb}
    \rho(\theta)=\frac{1}{2\pi}
    \sigma_P(\theta)+\int_{-\infty}^\infty\frac{d\theta'}{2\pi} 
\varphi(\theta-\theta')\rho^{(o)}(\theta'),
\end{equation}
where $\varphi=-i \frac{d}{d\theta} \log S(\theta)$ and
$\sigma_P(\theta)=dp/d\theta$. 
On the other hand, in the boundary case we have $\theta>0$ and the
densities satisfy the constraint
\begin{equation}
\label{Lieb-b}
    \rho(\theta)=\frac{1}{2\pi}
    \sigma_{ab}(\theta)+\int_{0}^\infty\frac{d\theta'}{2\pi} 
\big(\varphi(\theta-\theta')+\varphi(\theta+\theta')\big)\rho^{(o)}(\theta')
\end{equation}
where now
\begin{equation}
\label{sigma-ab}
  \sigma_{ab}(\theta)=\frac{d}{d\theta}\Big(2p(\theta)-i\frac{1}{L}
 \log R_{ab}(\theta)\Big)-2\pi \delta(\theta)
\end{equation}
where $R_{ab}(\theta)=R_a(\theta)R_b(\theta)S(-2\theta)$ and the extra
term containing $\delta(\theta)$ was introduced to cancel the
unphysical solutions corresponding to $\theta=0$.

It is known, that the the distribution of roots is given in both cases
by the thermodynamic Bethe Ansatz equations
\begin{equation}
\label{TBA}
  T \eps(\theta)=e(\theta) -T
  \int_{-\infty}^\infty\frac{d\theta'}{2\pi} 
\varphi(\theta-\theta') \log(1+e^{-\eps(\theta')}),
\end{equation}
where we introduced the pseudo-energy as
\begin{equation}
\label{epsdef}
  \frac{\rho(\theta)}{\rho^{(o)}(\theta)}=1+e^{\eps(\theta)}
\end{equation}

\subsection{The thermodynamic limit of the densities $\rho_N$ and $\bar\rho_N$}


The calculations will be based
on the techniques developed in the Algebraic Bethe Ansatz literature,
see for example Appendix A in \cite{korepin-LL1}. 


\medskip

In the periodic case the matrix elements of $\mathcal{J}$ read
\begin{equation*}
\MJ_{ik}=\delta_{ik}\Big(L\sigma_P(\theta_i)+\sum_{j=1}^{N}
\varphi(\theta_i-\theta_j)\Big) -\varphi(\theta_i-\theta_j) 
\end{equation*}
The matrix  $\mathcal{J}$ can be written as the product
\begin{equation*}
  \mathcal{J}=G\Theta,\qquad\text{where}
\end{equation*}
\begin{equation*}
  \Theta_{ij}=\delta_{ij} \gamma_j,\qquad
  G_{ij}=\delta_{ij}-\frac{\varphi(\theta_{ij})}{\gamma_j}\quad
\text{ and}
 \end{equation*}
\begin{equation}
\label{vtheloszor}
 \gamma_j=L\sigma_P(\theta_j)+\sum_{i=1}^{N} \varphi(\theta_{ij})
\end{equation}
With this notation 
\begin{equation*}
\rho_{N}=\det G_N \det \Theta_N
\end{equation*}
In the $L\to\infty$ limit we have from \eqref{Lieb}
\begin{equation*}
  \gamma_j\quad \to \quad 2\pi L\rho(\theta_j)
\end{equation*}
 The elements of $G_N$ can be written asymptotically as
\begin{equation*}
  G_{ij}=\delta_{ij}-\frac{1}{2\pi L}\frac{\varphi(\theta_{ij})}{\rho(\theta_j)}
\end{equation*}
Using \eqref{epsdef} we conclude that the limit of $\det G_{ij}$ is
given by the Fredholm determinant 
\begin{equation}
\label{ArminvanBuuren}
  \det\Big(\hat 1-\hat P^-\Big)
\end{equation}
where $\hat P^-$ is an integral operator acting as
\begin{equation}
\label{P}
  \big(\hat P^-
  (f)\big)(x)=\int_{-\infty}^\infty\frac{dy}{2\pi}\varphi(x-y)  \frac{1}{1+e^{\eps(y)}} f(y)
\end{equation}
Therefore
\begin{equation}
\label{ButchCassidy}
  \rho_N\quad \Rightarrow\quad    \det\Big(\hat 1-\hat P^-\Big)\times
  \prod_{j=1}^N 2\pi L\rho(\theta_j)
\end{equation}
The Fredholm determinant \eqref{ArminvanBuuren} is well-defined if the
operator $\hat P^-$ is trace class. This can be checked explicitly in
non-relativistic situations \cite{korepinBook}, or
in massive relativistic models at low temperatures, where 
\begin{equation*}
  \text{Tr} \Big(\big(\hat P^-\big)^\dagger \hat P^-\Big)\quad \sim\quad \ordo(e^{-2m/T}).
\end{equation*}
Here and in the rest of the paper we will assume that the Fredholm
determinants we encounter are always well-defined.  Even if this is not the case,
they could be made regular by introducing a rapidity
cut-off. This cut-off would not affect the final results, because
the $\ordo(1)$ pieces of the free energy will be
expressed as integral series which are well-defined for arbitrary
temperatures. 

\bigskip

Now we turn to the calculation of $\bar\rho_N$. In this case the
matrix elements of the Jacobian read
\begin{equation*}
  \bar{\MJ}_{ik}=\delta_{ik}\Big(L\sigma_{ab}(\theta_i)+\sum_{j=1}^{N}
\big(\varphi(\theta_i-\theta_j)+\varphi(\theta_i+\theta_j)\big)\Big)
 -\Big(\varphi(\theta_i-\theta_j)-\varphi(\theta_i+\theta_j) \Big)
\end{equation*}
The matrix can be written as
\begin{equation*}
  \bar{\mathcal{J}}=\bar G \bar\Theta,\qquad\text{with}
\end{equation*}
\begin{equation*}
  \bar\Theta_{ij}=\delta_{ij} \bar\gamma_j,\qquad\qquad
  \bar G_{ij}=\delta_{ij}-\frac{\varphi(\theta_i-\theta_j)-\varphi(\theta_i+\theta_j)}{\gamma_j},
 \end{equation*}
where now
\begin{equation}
\label{vth2}
 \gamma_j=L\sigma_{ab}(\theta_j)+\sum_{i=1}^{N} \big(
\varphi(\theta_i-\theta_j)+\varphi(\theta_i+\theta_j)\big)
\end{equation}
The quantities $\gamma_j$ have the same thermodynamic limit as in
the periodic case:
\begin{equation*}
  \gamma_j\quad \to \quad 2\pi L\rho(\theta_j)
\end{equation*}
In the thermodynamic limit one finds 
\begin{equation}
\label{Sundance-kid}
\bar  \rho_N\quad \Rightarrow\quad    \det\Big(\hat 1-\hat Q^-\Big)\times
  \prod_{j=1}^N 2\pi L\rho(\theta_j),
\end{equation}
where $\hat Q^-$ acts on functions defined on $\valos^+$ as
\begin{equation}
\label{Q-}
  \big(\hat Q^-
  (f)\big)(x)=\int_{0}^\infty\frac{dy}{2\pi}
\big(\varphi(x-y)-\varphi(x+y)\big)   \frac{1}{1+e^{\eps(y)}} f(y)
\end{equation}
Notice that the leading parts of \eqref{ButchCassidy} and
\eqref{Sundance-kid} have the same form, however the pre-factors are
different. This will play an important role in the next subsection.

\subsection{A heuristic derivation of the $g$-function (massive case)}

\label{RolfMaierBode}

Here we revisit the calculation of \cite{LeClair:1995uf} to determine the
exact $g$-function in massive relativistic boundary QFT.  We obtain
 obtain the correct boundary
independent part of the $g$-function using a simple heuristic argument.

In section 4 of \cite{LeClair:1995uf} the authors start with the
quantization conditions for $N$ particles \footnote{In
  \cite{LeClair:1995uf} (and also in \cite{Dorey:2004xk}) the roles of $L$ and $R$ are switched as
  opposed to our conventions}
\begin{equation}
  \label{BA-R}
  e^{i2mL\sinh\theta_j} R_a(\theta_j)R_b(\theta_j) \prod_{k\ne j}
  S(\theta_j-\theta_k) S(\theta_j+\theta_k)=1,
\qquad j=1\dots N,\quad\theta_j>0
\end{equation}
This can be written alternatively
as a quantization condition for $2n$ particles:
\begin{equation}
  \label{BA-R2}
  e^{i2mL\sinh\theta_j} R(\theta_i) \prod_{k\ne j}
  S(\theta_j-\theta_k) =1,
\qquad j=1\dots 2n
\end{equation}
with the additional constraint 
\begin{equation}
\label{constraint}
  \theta_{2n-j}=-\theta_j
\end{equation}
The function $R(\theta)$ in \eqref{BA-R2} is defined as
\begin{equation*}
  R(\theta)=R_a(\theta)R_b(\theta)S(-2\theta),
\end{equation*}

In \cite{LeClair:1995uf} the authors use \eqref{BA-R2} as a starting
point to derive the thermodynamics. They show that  \eqref{BA-R2}
yields the usual periodic-boundary-conditions TBA and  derive a
single $\mathcal{O}(1)$ correction to the free energy:
\begin{equation}
\label{wrong}
  \log (g_ag_b)=\frac{1}{4\pi} \int_{-\infty}^\infty d\theta\ \Theta_{ab}(\theta)
\log\big(1+e^{-\eps(\theta)}\big) \qquad\text{(Incomplete)}
\end{equation}
where
\begin{equation*}
  \Theta_{ab}(\theta)=-i\frac{d}{d\theta}\log R(\theta)-2\pi\delta(\theta)
\end{equation*}
where the term $\delta(\theta)$ is introduced to cancel the unphysical
states which do not respect the Pauli-principle. This result was
shown to be incomplete in \cite{Dorey:1999cj,Dorey:2004xk}.

We believe that the derivation in \cite{LeClair:1995uf} is correct in every
respect but one. There it was assumed that the change from system \eqref{BA-R}
to \eqref{BA-R2}-\eqref{constraint} is trivial. However, 
the previous subsection shows that the density of states of the two BA
systems behaves differently in the thermodynamic limit.
In fact, the periodic system with $2N$ particles has a density
\begin{equation*}
  \rho_{2N}\quad \sim \quad  \det\Big(\hat 1-\hat P^-\Big)\times
  \prod_{j=1}^{2N} 2\pi L\rho(\theta_j)
\end{equation*}
whereas the two copies of the boundary systems with $N$ particles have a
density
\begin{equation*}
  \Big(\bar\rho_N\Big)^2\quad \sim\quad {\det}^2 \Big(\hat 1-\hat Q^-\Big) 
\times  \prod_{j=1}^{2N} 2\pi L\rho(\theta_j)
\end{equation*}
In other words, there is a finite ratio between the (thermodynamic
dimensional) densities:
\begin{equation}
\label{ratio}
  \frac{\big(\bar\rho_N\big)^2}{\rho_{2N}}\quad\sim\quad 
\frac{\det^2\Big(\hat 1-\hat Q^-\Big) }{\det\Big(\hat 1-\hat P^-\Big) }
\end{equation}
This factor
does not grow with the volume $L$, therefore it does not affect the
thermodynamic limit of the distribution of roots. This means that
passing from the system \eqref{BA-R}
to \eqref{BA-R2}-\eqref{constraint} the
saddle point will not be changed and the 
distribution of roots will be described indeed by the
periodic-boundary-conditions TBA. However, a finite factor for the
density of states has to be kept in order to have a direct comparison
between the partition functions. In other words, the
ratio \eqref{ratio} has to be included in the overall normalization of the
partition function of the boundary system.

It is useful to explicitly evaluate this finite factor.
We start by employing the identity
\begin{equation}
\label{logdet}
  \det\Big(\hat 1-\hat K\Big)=
\exp\left\{-\sum_{n=1}^\infty \frac{1}{n}\text{Tr} K^n\right\}
\end{equation}
which is valid for any Fredholm-determinant.
In the present case we have
\begin{equation}
\label{present-case}
\frac{\big(\bar\rho_N\big)^2}{\rho_{2N}} \quad\Rightarrow\quad
\MN=\frac{\det^2\Big(\hat 1-\hat Q^-\Big) }{\det\Big(\hat 1-\hat P^-\Big) }=
\exp\left\{\sum_{n=1}^\infty  \frac{1}{n} \Big(2\text{Tr} {(P^-)}^n-\text{Tr} {(Q^-)}^n\Big)\right\}
\end{equation}
Notice that the kernels $P^-$ and $Q^-$ have different support. In can be
checked order by order, that the above formula results in
\begin{equation}
\label{vegso1}
  \log\MN=
\sum_{n=1}^\infty \frac{1}{n}
\int_{-\infty}^\infty \frac{d\theta_1}{2\pi}\dots \int_{-\infty}^\infty \frac{d\theta_n}{2\pi}
\left(\prod_{i=1}^n \frac{1}{1+e^{\eps(\theta_i)}}\right)
\varphi(\theta_1+\theta_2)\varphi(\theta_2-\theta_3)\dots \varphi(\theta_n-\theta_1)
\end{equation}
Adding this constant to  \eqref{wrong} one finds
\begin{equation}
\label{g-vegso1}
    \log (g_a)=\frac{1}{4\pi} \int_{-\infty}^\infty d\theta\ \Theta_{aa}(\theta)
\log\big(1+e^{-\eps(\theta)}\big) +\frac{1}{2}\log \MN
\end{equation}
where 
\begin{equation*}
  \Theta_{aa}(\theta)=\varphi_a(\theta)-\varphi(2\theta)
-\pi\delta(\theta),\quad\quad \varphi_a(\theta)=
-i\frac{d}{d\theta}\log R_a(\theta)
\end{equation*}
 This is the
exact non-perturbative $g$-function as obtained in
\cite{Dorey:2004xk}. We wish to note that the equivalence of
\eqref{present-case} and \eqref{vegso1} 
was already observed in \cite{Dorey:2005ak}.

This derivation does not take into
account the fluctuations around the saddle point, which are expected
to give a non-extensive contribution \cite{woynarovich}. 
However, equation
\eqref{BA-R2} already looks  like 
a periodic bc. quantization (with an additional factor) and it is generally
accepted, that there is no $\ordo(1)$ piece in the periodic case.
This implies that eq. \eqref{g-vegso1} already takes into account all
$\ordo(1)$ contributions, ie. there are no additional terms due
to the fluctuations.

The reader may be concerned that our arguments are not convincing and
that the proposed
re-normalization 
might be dictated by the result itself. We agree that our derivation
of 
\eqref{g-vegso1} is certainly not rigorous, however, it already shows that
the density of states play an important role in the definition of the
partition function. 
In the next section we present a general
calculation built on first principles, which properly takes into 
account both the density of states and the fluctuations around the
saddle point. In the massive case we obtain
the same result \eqref{g-vegso1}.

\section{$\ordo(1)$ pieces to the free energy -- formal derivation}

\label{fasza}

In this section we present a general framework to determine the
$\ordo(1)$ pieces to the free energy. We consider Bethe-Ansatz systems with one particle type
and one rapidity variable $\theta$ which takes its 
values from a domain $\mathcal{B}$. The Bethe-Yang equations are
assumed to take the form
\begin{equation}
  \label{generalBA}
e^{iQ_j(\theta_1,\dots,\theta_N)}\equiv 
e^{i\alpha p_jL} R_{BY}(\theta_j)  \prod_{k\ne j} \mathcal{S}_{BY}(\theta_j,\theta_k)=1
\end{equation}
where $p_j=p(\theta_j)$. The pure number $\alpha$ and the
phases $R_{BY}(\theta_j)$ and $\mathcal{S}_{BY}(\theta_j,\theta_k)$ depend on the 
problem at hand. In massive relativistic models (or
arbitrary non-relativistic models) with periodic boundary conditions we have
\begin{equation*}
\begin{split}
  \alpha=1\quad\quad R_{BY}(\theta)=1 \\
  \mathcal{S}_{BY}(\theta_j,\theta_k)=S(\theta_j-\theta_k)
\end{split}
\end{equation*}
In the case of integrable integrable boundaries 
\begin{equation*}
\begin{split}
  \alpha=2\quad\quad R_{BY}(\theta)=R_a(\theta)R_b(\theta) \\
 \mathcal{S}_{BY}(\theta_j,\theta_k)=S(\theta_j-\theta_k)
S(\theta_j+\theta_k)
\end{split}
\end{equation*}
For the treatment of massless relativistic models see section \ref{massless}.
It is assumed that the unitarity condition is satisfied in the form
\begin{equation*}
   \mathcal{S}_{BY}(\theta_j,\theta_k) \mathcal{S}_{BY}(-\theta_j,-\theta_k)=1
\end{equation*}
To investigate the thermodynamic limit, we split up the $\theta$ axis
into intervals $\Delta \theta$ and we use 
the densities $\rho_o(\theta)$ and $\rho_h(\theta)$ for the occupied
states and the holes. The total density of states is $\rho(\theta)=\rho_o(\theta)+\rho_h(\theta)$.
It follows from \eqref{generalBA} that the densities satisfy the
constraint
\begin{equation}
  \label{Lieb2}
    \rho_h(\theta)+ \rho_o(\theta)=\frac{1}{2\pi}
    \sigma(\theta)+\int_{\mathcal{B}}\frac{d\theta'}{2\pi} 
K_1(\theta,\theta')\rho^{(o)}(\theta'),
\end{equation}
where 
\begin{equation*}
  K_1(\theta,\theta')=-i \frac{\partial}{\partial \theta} \log \mathcal{S}_{BY}(\theta,\theta')
\end{equation*}
and
\begin{equation}
\label{general-sigma}
   \sigma(\theta)=\alpha \frac{d}{d\theta} p(\theta)+\Theta(\theta)
 \end{equation}
with
\begin{equation}
\label{THab}
  \Theta(\theta)=-i\frac{d}{d\theta}\frac{1}{L}
 \log \Big(R_{BY}(\theta)\mathcal{S}_{BY}(-\theta,-\theta) \Big)
\end{equation}
Depending on the problem at hand there can be an extra $-2\pi
\delta(\theta)$ term to \eqref{THab} to take into account the Pauli-principle.

Based on the calculations of the previous section we assume that the
density of states for an $N$-particle state can be written as
\begin{equation}
\label{density-assumption}
  \rho_N(\theta_1,\dots,\theta_N)=\MD \times \prod_{j=1}^N \big(2\pi L
  \rho(\theta_j)\big)
\end{equation}
where $\MD$ is a bounded
finite number  for a smooth distribution of roots. In the
thermodynamic limit it is
given typically by a Fredholm-determinant; the explicit calculation
for a generic case will be presented later in this section.

The partition function $Z$ is obtained as a summation over all possible
momentum quantum numbers in a finite volume $L$. In the thermodynamic
limit it is evaluated by a functional integral over all
possible density functions. 
The usual
prescription to define a regularized functional integral for $Z$ is
\begin{equation}
\label{rosszF}
  \sum_{configurations}\quad \Rightarrow \quad \int\dots\int \prod_\theta
  \big(L\Delta \theta d\rho_o(\theta)\big)
\end{equation}
Here the product is over the discretization points and there is one integral for the density at
that point. 

We believe that this prescription is not correct for Bethe
Ansatz systems. The partition function depends on the configuration
which minimizes the free energy and also on the {\it number of available states} for
that particular configuration. The calculations which result in
formulas like \eqref{density-assumption} show that the number of
available states for a Bethe Ansatz system does not behave like the
density in a free
theory. Instead, there is the non-trivial finite factor $\MD$ which
has to be taken into account. In other words, 
the constraints for the configuration space of the theory have to be incorporated in the
definition of the functional integral.

In the usual evaluation schemes the entropy considerations are worked
out in rapidity space, and the functional integral is performed using
\eqref{rosszF}. In this procedure one is
free to choose the discretization mesh $\Delta \theta$ in a fairly
wide range (for a discussion on this point see
\cite{zam-tba,woynarovich}). The calculations are safe in the sense
that the end result does not depend on the actual value of $\Delta
\theta$. However, $\Delta\theta$ should be chosen large enough so that
there are enough particles in the interval
$(\theta,\theta+\Delta\theta)$ to make the entropy considerations meaningful.
On the other hand, the ``ultimate discretization''  would be
to have one discretization point for each particle in the system; in
this case the
variation of the free energy would be performed on the level of
the momentum quantum numbers. Such a scheme is obviously not amenable
for practical purposes. The solution is to perform a transformation from the
space of momentum quantum numbers to the space of rapidities; the
important point is to
keep the Jacobian associated to this mapping. In the thermodynamic
limit the Jacobian behaves as given by  \eqref{density-assumption},
where the product over $j=1\dots N$ can be interpreted as the ``ultimate
discretization'' of the functional integral \eqref{rosszF}. A very
important observation is that in the thermodynamic limit the pre-factor $\mathcal{N}$ does not
depend on the number of particles involved, therefore we conclude that
it must be present in any discretization scheme.

Based on the above considerations we propose the following
definition of the functional integral:
\begin{equation}
\label{Felozetes}
  \sum_{configurations}\quad  \Rightarrow\quad \int\dots\int  \MD
\prod_\theta  \big(L\Delta \theta d\rho_o(\theta)\big)
\end{equation}
The factor $\MD$ depends on the
particular thermodynamic configuration which minimizes the free energy
functional. However, experience shows that $\MD$ is a finite
number which does not grow with the volume. Therefore it is expected
that it will not shift the position of the saddle point and its only
effect is to correctly normalize the partition function. It follows that
the prescription \eqref{Felozetes} is equivalent to
\begin{equation}
\label{F}
  \sum_{configurations}\quad  \Rightarrow\quad \MD \int\dots\int \prod_\theta
  \big(L\Delta \theta d\rho_o(\theta)\big)
\end{equation}
and the difference between \eqref{Felozetes} and \eqref{F} vanishes
in the thermodynamic limit.

\bigskip

In the following we evaluate the partition function using the
prescription \eqref{F}. The calculation proceeds in four steps:
\begin{enumerate}
\item Establishing the Thermodynamic Bethe Ansatz using the usual
  prescription \eqref{rosszF}
\item Evaluating the contribution of the quadratic fluctuations around
  the saddle point
\item Evaluating the normalization factor $\mathcal{N}$ 
\item Writing down the final result with the correct normalization \eqref{F}
\end{enumerate}

The treatment of the quadratic fluctuations 
follows exactly the same way as in
\cite{woynarovich}; we simply restate the results of
\cite{woynarovich} in our general framework.
In order the keep the exposition simple we omit
some of the technical details and the checks of the various
approximations we make. For a careful treatment the reader is referred to
the original work \cite{woynarovich}.

\newpage

\bigskip

{\bf Step 1}

\bigskip

We evaluate the partition function using the functional integral
\eqref{rosszF};
the entropy considerations are worked out in rapidity space.
The number of micro-canonical configurations is given by 
\begin{equation*}
  \Omega= \prod_\theta \omega(\rho_o(\theta))
\end{equation*}
where
\begin{equation*}
  \omega=
  \begin{pmatrix}
    L\Delta \theta \rho(\theta)\\   L\Delta \theta \rho_o(\theta)
  \end{pmatrix}=
\frac{\big(L\Delta \theta \rho(\theta)\big)!}
{\big(L\Delta \theta \rho_o(\theta)\big)!\big(L\Delta \theta \rho_h(\theta)\big)!}
\end{equation*}
We approximate it using Stirling-formula as
\begin{equation*}
\log  \omega=L\Delta \theta
s(\rho(\theta))+\varsigma(\rho(\theta))+\dots\ ,
\end{equation*}
where
\begin{equation*}
  s(\rho(\theta))=\rho(\theta)\log(\rho(\theta))
-\rho_o(\theta)\log(\rho_o(\theta))-\rho_h(\theta)\log(\rho_h(\theta))
\end{equation*}
and
\begin{equation*}
\varsigma(\rho(\theta))=-\frac{1}{2}\log(2\pi L\Delta\theta)+
\frac{1}{2}\log\frac{\rho(\theta)}{\rho_o(\theta)\rho_h(\theta)}  
\end{equation*}
The free energy functional 
 can be written as
\begin{equation}
\label{Ffunct}
  F[\rho(\theta)]=L\sum_\theta \Big(e(\theta)\rho_o(\theta)-Ts(\theta)\Big)\Delta \theta
\end{equation}
The usual minimalization procedure yields the integral equation
\begin{equation}
\label{TBAg}
  e(\theta)/T=\eps(\theta)+\int_\MB \frac{d\theta'}{2\pi}\  K_1(\theta,\theta')
  \log\big(1+e^{-\eps(\theta')}\big),
\end{equation}
where we introduced the pseudo-energy function as
\begin{equation}
\label{epsdef2}
  \frac{\rho_o(\theta)}{\rho_h(\theta)}=e^{-\eps(\theta)}
\end{equation}
The ``minimal part'' of the free
energy can be expressed simply as
\begin{equation}
\label{Fmin}
  F_{\rm min}=-LT \int_\MB \frac{d\theta}{2\pi} \
  \sigma(\theta)\log\big(1+e^{-\eps(\theta)}\big),
\end{equation}
where $\sigma(\theta)$ is given by \eqref{general-sigma}.

\bigskip
{\bf Step 2}
\bigskip

We consider the fluctuations around the saddle point solution. As a
first step we expand the free
energy functional \eqref{Ffunct} around $\rho_o(k)$ and $\rho_h(k)$: 
\begin{equation}
F_L\left[\rho_o(\theta)\right]\simeq F_{\rm min}-T\sum_\theta L\Delta \theta {1\over2}
\left({\left(r_o(\theta)+r_h(\theta)\right)^2\over\rho_o(\theta)+\rho_h(\theta)}-
{r_h^2(\theta)\over\rho_h(\theta)}-
{r_o^2(\theta)\over\rho_o(\theta)}\right)\,.
\end{equation}
Here the quantities 
\begin{equation}
r_o(\theta)=\delta \rho_o(\theta)\quad{\rm and}\quad
r_h(\theta)=\delta \rho_h(\theta)
\end{equation}
are constrained by 
\begin{equation}
\label{rrhdBg}
r_o(\theta)+r_h(\theta)=\sum_{\theta^{\prime}}K_1\left(\theta,\theta^{\prime}\right)r_o(\theta^{\prime})
\Delta \theta^{\prime}\,.
\end{equation}
The absolute magnitude of the partition function is then evaluated as
\begin{eqnarray}
\label{funcintdBg}
&&Z=\MD\ \times\ e^{-\beta F_{\rm min}}\ \times \
\int\cdots\int\prod_\theta\left(L\Delta \theta dr(\theta)\right)\times\\
&&\exp\left\{-\sum_\theta\left(L\Delta \theta{1\over2}
\left({\left(r_o(\theta)+r_h(\theta)\right)^2\over\rho_o(\theta)+\rho_h(\theta)}-
{r_h^2(\theta)\over\rho_h(\theta)}-
{r_o^2(\theta)\over\rho_o(\theta)}\right)-
{1\over2}\ln{\rho_{0}+\rho_h\over
\rho_{0}\rho_hL\Delta \theta2\pi}\right)\right\}\,,\nn
\end{eqnarray}
We introduce new integration variables
$\xi(\theta)$ according to
\begin{equation}
  \xi(\theta)=\sqrt{L\Delta\theta \frac{1}{2}\frac{\rho(\theta)}
{\rho_o(\theta)\rho_h(\theta)}} r(\theta)
\end{equation}
This leads to
\begin{equation}
-\sum_{\theta,\theta^{\pr},\theta^{\dpr}}\xi(\theta)\left(\delta_{\theta^{\pr},\theta}-M_{\theta^{\pr},\theta}\right)
\left(\delta_{\theta^{\pr},\theta^{\dpr}}-M_{\theta^{\pr},\theta^{\dpr}}\right)\xi(\theta^{\dpr})\,,
\end{equation}
where $\delta_{\theta,\theta^{\pr}}$ is the Kronecker symbol, and
\begin{equation}
M_{\theta,\theta^{\pr}}={1\over\rho_h(\theta)}\sqrt{{\rho_o(\theta)\rho_h(\theta)
\Delta \theta\over\rho_o(\theta)+\rho_h(\theta)}}K_1(\theta,\theta^{\pr})
\sqrt{{\rho_o(\theta^{\pr})\rho_h(\theta^{\pr})
\Delta \theta^{\pr}\over\rho_o(\theta^{\pr})+\rho_h(\theta^{\pr})}}
\end{equation}
Finally, changing the integration variable to $\xi$ in (\ref{funcintdBg})
we have
\begin{eqnarray}
Z&=&\MD e^{-\beta F_{\rm min}}
\int\cdots\int\prod_\theta\left({d\xi(\theta)\over\pi}\right)
\exp\left\{-\sum_{\theta,\theta^{\pr},\theta^{\dpr}}\xi(\theta)\left(\delta_{\theta,\theta^{\pr}}-M_{\theta,\theta^{\pr}}\right)
\left(\delta_{\theta^{\dpr},\theta^{\pr}}-M_{\theta^{\dpr},\theta^{\pr}}\right)\xi(\theta^{\dpr})\right\}\nn\\
&&\\
&=&\MD e^{-\beta F_{\rm min}}
\left(\det\left[\delta_{\theta,\theta^{\pr}}-K_{\theta,\theta^{\pr}}\right]\right)^{-1}\,
\nn
\end{eqnarray}
with
\begin{equation}
K_{\theta,\theta^{\pr}}={\sqrt{\rho_h(\theta)}}\cdot M_{\theta,\theta^{\pr}}\cdot
{1\over\sqrt{\rho_h(\theta^{\pr})}}=\sqrt{{\rho_o(\theta)
\Delta \theta\over\rho_o(\theta)+\rho_h(\theta)}}K_1(\theta,\theta^{\pr})
\sqrt{{\rho_o(\theta^{\pr})
\Delta \theta^{\pr}\over\rho_o(\theta^{\pr})+\rho_h(\theta^{\pr})}}\,.
\end{equation}
Taking the limit $\sum_\theta \Delta\theta\to\int_\MB d\theta$  yields 
\begin{equation}
\label{joZ0}
  Z= \mathcal{N}e^{-\beta F_{\rm min}} \Big(\det\big(\hat 1-\hat K_1\big)\Big)^{-1}
\end{equation}

\bigskip
{\bf Step 3}
\bigskip

The density of states is given by the determinant
\begin{equation*}
  \rho_N(\theta_1,\dots,\theta_N)=\det \mathcal{J},\qquad
  \mathcal{J}_{ik}=\frac{\partial Q_i}{\partial \theta_k}
\end{equation*}
A straightforward calculation shows that the matrix elements are given by
\begin{equation*}
  {\MJ}_{ik}=\delta_{ik}\Big(L\sigma(\theta_i)+\sum_{j=1}^{N} K_1(\theta_i,\theta_j)
\Big)-K_2(\theta_i,\theta_j),
\end{equation*}
where 
\begin{equation*}
 K_2(\theta,\theta')=i \frac{\partial}{\partial \theta'} \log
 \mathcal{S}_{BY}(\theta,\theta')
\end{equation*}
The matrix can be written as
\begin{equation*}
  {\mathcal{J}}= G \Theta,\qquad\text{where now}
\end{equation*}
\begin{equation*}
  \Theta_{ij}=\delta_{ij} \gamma_j,\qquad\qquad
   G_{ij}=\delta_{ij}-\frac{K_2(\theta_i,\theta_j)}{\gamma_j}
 \end{equation*}
with
\begin{equation}
\label{vth3}
 \gamma_j=L\sigma(\theta_j)+\sum_{i=1}^{N} K_1(\theta_1,\theta_2)
\end{equation}
It follows from \eqref{Lieb2} that in the thermodynamic limit
\begin{equation*}
  \gamma_j\quad \to \quad 2\pi L\rho(\theta_j)
\end{equation*}
The elements of $\bar G_N$ can be written asymptotically as
\begin{equation*}
 G_{ij}=\delta_{ij}-\frac{1}{2\pi   L}
\frac{K_2(\theta_i,\theta_j)}
{\rho(\theta_j)}
\end{equation*}
Using \eqref{epsdef2} we conclude that the limit of $\det G_{ij}$ is
given by the Fredholm determinant with kernel
$K_2(\theta,\theta')$. Therefore the normalization constant of the
free energy functional \eqref{F} is given by
\begin{equation}
\label{NNN}
  \mathcal{N}=\det\Big(\hat 1-\hat K_2\Big)
\end{equation}

\bigskip
{\bf Step 4 -- Our main result}
\bigskip

Substituting \eqref{NNN} into \eqref{joZ0} yields
\begin{equation}
\label{joZ}
  Z= e^{-\beta F_{\rm min}} \frac{\det\big(\hat 1-\hat K_2\big)}{\det\big(\hat 1-\hat K_1\big)}
\end{equation}
Here $\hat K_1$ and $\hat K_2$ are integral operators which act on functions defined on
$\mathcal{B}$ as
\begin{equation*}
  \big(\hat K_j
  (f)\big)(x)=\int_{\mathcal{B}}\frac{dy}{2\pi}K_j(x,y)
  \frac{1}{1+e^{\eps(y)}} f(y)\qquad j=1,2
\end{equation*}
and the kernels are given by
\begin{equation*}
 K_1(\theta,\theta')=-i \frac{\partial}{\partial \theta} \log
 \mathcal{S}_{BY}(\theta,\theta')
\qquad\quad
 K_2(\theta,\theta')=i \frac{\partial}{\partial \theta'} \log
 \mathcal{S}_{BY}(\theta,\theta')
\end{equation*}
The phase shift $\mathcal{S}_{BY}(\theta,\theta')$ is defined implicitly
by the Bethe-Yang equations \eqref{generalBA} and $F_{\rm min}$ is
given by formula \eqref{Fmin}.


\section{Explicit examples -- massive relativistic models}

In this section we evaluate \eqref{joZ} explicitly in four different cases; we
restrict ourselves to relativistic models. Non-relativistic ones 
can be treated in 
the same manner. 

\subsection{Free fermionic gas}

In this case 
 \begin{equation*}
    \MB=\valos\quad \text{and}\quad \sigma(\theta)=m\cosh\theta.
  \end{equation*}
There is no interaction between the particles, therefore
  \begin{equation*}
      K_1(\theta,\theta')=K_2(\theta,\theta')=0 \quad \text{and}\quad  \MD=1.
  \end{equation*}
The thermodynamics is trivial:
\begin{equation*}
  \eps(\theta)=e(\theta)/T=m\cosh\theta/T
\end{equation*}
and
\begin{equation*}
  F_{\rm min}=-LT \int_{-\infty}^\infty \frac{d\theta}{2\pi} \
m  \cosh\theta \log\big(1+e^{-m\cosh\theta/T}\big)
\end{equation*}
There is no $\ordo(1)$ piece.

\subsection{Interacting particles with periodic boundary conditions}

In this case
  \begin{equation*}
    \MB=\valos\quad \text{and}\quad \sigma(\theta)=m\cosh\theta.
  \end{equation*}
The scattering phase shift is
\begin{equation*}
  \mathcal{S}_{BY}(\theta,\theta')=S(\theta-\theta').
\end{equation*}
Therefore the two integral kernels are given by
  \begin{equation*}
      K_1(\theta,\theta')=K_2(\theta,\theta')=\varphi(\theta-\theta') 
  \end{equation*}
The free energy reads
\begin{equation}
\label{Fpbc}
  F_{\rm min}^{p}=-LT \int_{-\infty}^\infty \frac{d\theta}{2\pi} \
m  \cosh\theta \log\big(1+e^{-\eps(\theta)}\big)
\end{equation}
Equation \eqref{joZ} results in
\begin{equation*}
    Z=\frac{\det\big(\hat 1-\hat K_2\big)}{\det\big(\hat 1-\hat
      K_1\big)}
 e^{-\beta F_{\rm min}^p} 
=e^{-\beta F_{\rm min}^p}
\end{equation*}
The two Fredholm-determinants 
coincide and there is no $\ordo(1)$ piece.

\subsection{Periodic bc. with a purely transmitting defect}

One can consider the Bethe Ansatz equations in the presence of a
purely transmitting defect:
\begin{equation*}
  e^{ip_jL}T(\theta_j)\prod_{k\ne j}S(\theta_j-\theta_k)=1,
\end{equation*}
where $T(\theta)$ describes the scattering between the particles and the
defect. Similar to the previous case one has
  \begin{equation*}
      K_1(\theta,\theta')=K_2(\theta,\theta')=\varphi(\theta-\theta') 
  \end{equation*}
However, the function $\sigma(\theta)$ is now given by
\begin{equation*}
  \sigma(\theta)=m\cosh\theta+\varphi_T(\theta),\qquad\text{where}\qquad
\varphi_T(\theta)=-i\frac{1}{L}\frac{d}{d\theta} \log  T(\theta)
\end{equation*}
The partition function is expressed as
\begin{equation*}
     Z=\frac{\det\big(\hat 1-\hat K_2\big)}{\det\big(\hat 1-\hat
      K_1\big)}
 e^{-\beta F_{\rm min}^p-\beta F_T} 
=e^{-\beta F_{\rm min}^p-\beta F_T}
\end{equation*}
where $F_T$ is an $\ordo(1)$ piece given by
\begin{equation*}
  F_T=-T \int_{-\infty}^\infty 
 \frac{d\theta}{2\pi} \
\varphi_T(\theta) \log\big(1+e^{-\eps(\theta)}\big)
\end{equation*}
This is in agreement with the previous results in the literature
(see for example \cite{martins-1994}). 

\subsection{Integrable boundaries}

Here the domain of integrations is $\MB=\valos^+$ and 
the scattering phase shift is given by
\begin{equation*}
    \mathcal{S}_{BY}(\theta,\theta')=S(\theta-\theta')S(\theta+\theta')
\end{equation*}
The integral kernels are given by
\begin{equation*}
\begin{split}
   K_1(\theta,\theta')&=\varphi(\theta-\theta')+\varphi(\theta+\theta')\\
  K_2(\theta,\theta')&=\varphi(\theta-\theta')-\varphi(\theta+\theta')
\end{split}
\end{equation*}
The associated Fredholm-determinants are $\det\big(\hat 1-\hat Q^\pm\big)$
where the operators $\hat Q^\pm$ are defined as
\begin{equation}
\label{Q+-}
  \big(\hat Q^\pm
  (f)\big)(x)=\int_{0}^\infty\frac{dy}{2\pi}
\big(\varphi(x-y)\pm\varphi(x+y)\big)   \frac{1}{1+e^{\eps(y)}} f(y)
\end{equation}
The function $\sigma(\theta)$ is given by
\begin{equation*}
   \sigma(\theta)=2m\cosh\theta+\Theta_{ab}(\theta)
 \end{equation*}
where 
\begin{equation*}
  \Theta_{ab}(\theta)=-i\frac{d}{d\theta}\frac{1}{L}
 \log R_{ab}(\theta)-2\pi \delta(\theta)
\end{equation*}
 The TBA equation \eqref{TBAg} takes the form
\begin{equation*}
 m\cosh\theta/T=\eps(\theta)+\int_0^\infty \frac{d\theta'}{2\pi}\ 
\big(\varphi(\theta-\theta')+\varphi(\theta+\theta') \big)
  \log\big(1+e^{-\eps(\theta')}\big)  
\end{equation*}
One can define the pseudo-energy also for negative values of
rapidities as $\eps(\theta)=\eps(-\theta)$, then the above equation above is equivalent to
\begin{equation*}
 m\cosh\theta/T=\eps(\theta)+\int_{-\infty}^\infty \frac{d\theta'}{2\pi}\ 
\varphi(\theta-\theta')
  \log\big(1+e^{-\eps(\theta')}\big),  
\end{equation*}
which is the usual periodic boundary conditions TBA. The minimal part
of the free energy
is expressed as
\begin{equation*}
  \begin{split}
     F_{\rm min}
&= -LT \int_0^\infty \frac{d\theta}{2\pi} \
     (2m\cosh\theta+\frac{1}{L}\Theta_{ab}(\theta))\log\big(1+e^{-\eps(\theta)}\big)=
F_{\rm min}^{\rm p}+F_{ab}
  \end{split}
\end{equation*}
where $F_{\rm min}^{\rm p}$ is the free energy of the periodic bc
system \eqref{Fpbc} and $F_{ab}$ is an $\ordo(1)$ piece given by
\begin{equation}
\label{Fab}
  F_{ab}=-T \int_{-\infty}^\infty \frac{d\theta}{4\pi} 
\Theta_{ab}(\theta)\log\big(1+e^{-\eps(\theta)}\big)
\end{equation}
The partition function reads
\begin{equation*}
  Z=\frac{\det\big(\hat 1-\hat Q^-\big)}{\det\big(\hat 1-\hat
    Q^+\big)}e^{-\beta F_{\rm min}^p -\beta F_{ab}} 
\end{equation*}
Therefore the $\ordo(1)$ piece to the free energy is given by
\begin{equation}
\label{g-vegso0}
\log(g_ag_b)=-\beta F_{ab}+\log  \frac{\det\big(\hat 1-\hat Q^-\big)}{\det\big(\hat 1-\hat
    Q^+\big)}
\end{equation}
In Appendix A we show that the above ratio of Fredholm-determinants
reproduces the constant $\log\mathcal{A}$ defined in
\eqref{vegso1}. Choosing the boundary conditions $a=b$ the final result can be written as
\begin{eqnarray}
  \label{g-vegso2}
&&\hspace{1cm}\log(g_a)=\frac{1}{2} \int_{-\infty}^\infty \frac{d\theta}{4\pi} 
\Theta_{aa}(\theta)\log\big(1+e^{-\eps(\theta)}\big)+\\
\nonumber
&&+\frac{1}{2}
\sum_{n=1}^\infty \frac{1}{n}
\int_{-\infty}^\infty \frac{d\theta_1}{2\pi}\dots \int_{-\infty}^\infty \frac{d\theta_n}{2\pi}
\left(\prod_{i=1}^n \frac{1}{1+e^{\eps(\theta_i)}}\right)
\varphi(\theta_1+\theta_2)\varphi(\theta_2-\theta_3)\dots \varphi(\theta_n-\theta_1)
\end{eqnarray}
 This is the
exact non-perturbative $g$-function as obtained in \cite{Dorey:2004xk}.

\section{Explicit examples -- massless relativistic models}

\label{massless}

In this section we consider massless relativistic theories with diagonal
scattering \cite{Zamolodchikov:1991vx,Fendley:1993xa}; 
for an introduction to massless scattering the reader is
referred to \cite{Fendley:1993jh}.
We assume that there is only one particle type in the spectrum.
There is an effective doubling because one has to treat
left-moving and right-moving particles separately.
We consider parity-invariant theories; the S-matrices are given by
\begin{equation*}
  S_{LL}(\theta)=S_{RR}(\theta) \qquad \qquad  S_{LR}(\theta)=S_{RL}(\theta) 
\end{equation*}
The TBA equations are usually written down for periodic boundary
conditions, in which case 
 the TBA is a two-component system:
\begin{equation}
  \begin{split}
    \eps_1(\theta)&=\frac{1}{2} mR e^\theta -\int_{-\infty}^\infty
    \frac{d\theta'}{2\pi}\Big(\varphi_{11}(\theta-\theta')
\log\big(1+e^{-\eps_1(\theta')}\big)
+\varphi_{12}(\theta-\theta') \log\big(1+e^{-\eps_2(\theta')}\big)\Big)\\
   \eps_2(\theta)&=\frac{1}{2} mR e^{-\theta} -\int_{-\infty}^\infty
    \frac{d\theta'}{2\pi}
\Big(\varphi_{21}(\theta-\theta')
\log\big(1+e^{-\eps_1(\theta')}\big)
+\varphi_{22}(\theta-\theta') \log\big(1+e^{-\eps_2(\theta')}\big)\Big)
  \end{split}
\end{equation}
The kernels are defined as 
\begin{equation*}
  \varphi_{11}(\theta)=\varphi_{22}(\theta)=-i \frac{d}{d\theta} \log
  S_{RR}(\theta)\qquad
 \varphi_{12}(\theta)=\varphi_{21}(\theta)=-i \frac{d}{d\theta} \log
  S_{LR}(\theta)
\end{equation*}
One can then use the symmetry $\eps_1(\theta)=\eps_2(-\theta)$ 
to transform the above system into a single equation:
\begin{equation}
\label{masslessTBA-akurvaanyad}
    \eps(\theta)=\frac{1}{2} mR e^\theta -\int_{-\infty}^\infty
    \frac{d\theta'}{2\pi}\big(\varphi_{11}(\theta-\theta')+\varphi_{12}(\theta+\theta')\big)
\log\big(1+e^{-\eps(\theta')}\big)
\end{equation}
The free energy is expressed as
\begin{equation}
\label{masslessFminA}
  F_{\rm min}^p=-LT \int_{-\infty}^\infty  \frac{d\theta}{2\pi} m e^\theta
\log\big(1+e^{-\eps(\theta)}\big)
\end{equation}
It is generally expected that for periodic boundary conditions there is no $\ordo(1)$ piece.

\bigskip

In the following we apply the formalism of section \ref{fasza} to determine the $g$-function in
the presence of two integrable boundaries with reflection factors
$R_{a}(\theta)$ and $R_{b}(\theta)$. Contrary to the periodic
case, in the open system there is no distinction between the
left-movers and the right-movers. When a left-mover scatters off the
left-boundary, it becomes a right-mover with the same energy and
reversed momentum. Therefore both the Bethe Ansatz and also the
thermodynamics can be written down in terms of only one particle
species and one pseudo-energy function. 

We use simple heuristic arguments to write down the Bethe-Yang equations.
The particles will
be parametrized with the rapidity variable $\theta\in\valos$ which
refers to the situation when the particle is moving to the right, ie.
\begin{equation*}
  e(\theta)=\frac{1}{2} me^{\theta}\qquad p(\theta)=\frac{1}{2}me^{\theta}
\end{equation*}
When a particle with rapidity $\theta$ is taken back and forth in the open system it meets
every other particle twice. 
The two scattering processes are described by $S_{RR}(\theta-\theta')$
and $S_{LR}(\theta+\theta')$. 
The Bethe-Yang equations read
\begin{equation}
\label{masslessBA2}
  e^{i2p_jL}R_a(\theta)R_b(\theta)\prod_{k\ne j}S_{RR}(\theta_k-\theta_j)S_{LR}(\theta_k+\theta_j)=1
\end{equation}
To establish the connection with our general formalism we extract the
quantities
\begin{equation*}
  \alpha=2, \qquad \qquad R_{BY}(\theta)=R_a(\theta)R_b(\theta)S_{LR}(-2\theta),
\end{equation*}
and for the scattering phase shift we find
\begin{equation*}
  \mathcal{S}_{BY}(\theta,\theta')=S_{RR}(\theta-\theta') S_{LR}(\theta+\theta')
\end{equation*}
which yields the integration kernels
\begin{equation*}
\begin{split}
  K_1(\theta,\theta')=\varphi_{11}(\theta-\theta')+\varphi_{12}(\theta+\theta')\\
  K_2(\theta,\theta')=\varphi_{11}(\theta-\theta')-\varphi_{12}(\theta+\theta')
\end{split}
\end{equation*}
The function $\sigma(\theta)$ is given by
\begin{equation*}
  \sigma(\theta)=me^\theta+\Theta_{ab}(\theta)
\end{equation*}
where 
\begin{equation*}
  \Theta_{ab}(\theta)=-i\frac{d}{d\theta}\frac{1}{L}
 \log R_{ab}(\theta),\quad\quad R_{ab}(\theta)=R_a(\theta)R_b(\theta)S_{LR}(-2\theta) 
\end{equation*}
Note that contrary to \eqref{sigma-ab} the $\delta(\theta)$ term is
missing from $\sigma(\theta)$ because there are no formal solutions with $\theta=0$ which
should be canceled. 

Formula \eqref{TBAg} yields the TBA equation
\begin{equation}
\label{ujTBA}
    \eps(\theta)=\frac{1}{2} mR e^\theta -\int_{-\infty}^\infty
    \frac{d\theta'}{2\pi}
\big(\varphi_{11}(\theta-\theta')+\varphi_{12}(\theta+\theta') \big)
\log\big(1+e^{-\eps(\theta')}\big)
\end{equation}
This equation coincides with \eqref{masslessTBA-akurvaanyad},
although it was derived from
a conceptually different Bethe Ansatz. The interpretation is straightforward:
the distribution of roots (and therefore the $\ordo(L)$
pieces of the free energy) do not depend on the boundary conditions,
as it is expected on general grounds.

Formula \eqref{Fmin} yields
\begin{equation*}
  F_{\rm min}=F_{\rm min}^p+F_{ab}
\end{equation*}
Here $F_{\rm min}^p$ is given by \eqref{masslessFminA} and $F_{ab}$ is
an $\ordo(1)$ piece given by
\begin{equation}
\label{masslessFab}
  F_{ab}=-T
\int_{-\infty}^\infty \frac{d\theta}{2\pi} 
\Theta_{ab}(\theta)\log\big(1+e^{-\eps(\theta)}\big)
\end{equation}
Notice the factor of 2 as compared to \eqref{Fab}.

Putting everything together, equation \eqref{joZ} yields
\begin{equation}
\label{mounteverest}
  Z=\frac{\det\big(1-\hat K_2\big)}{\det\big(1-\hat K_1\big)}e^{-\beta
    F_{\rm min}^p-\beta F_{ab}}
\end{equation}

In the next two subsections we explicitly work out the details
for simple scattering theories with one particle species. The
generalization to other models with more than one particles (for
example the scattering theory in \cite{Dorey:1996ms}) can be treated
with the straightforward extension of
\eqref{mounteverest}.  Finally we mention that
the boundary independent part of \eqref{mounteverest}  can be written in the form
\begin{eqnarray}
  \label{verygood}
&&\log  \frac{\det\big(1-\hat K_2\big)}{\det\big(1-\hat  K_1\big)}=
\sum_{n=1}^\infty \frac{1}{n}\Big(\text{Tr}  {(\hat K_1)}^n-\text{Tr}{(\hat K_2 )}^n\Big)=
\\
\nonumber
 && \sum_{n=1}^\infty \frac{1}{n} \sum_{a_1\dots a_n}
\int_{-\infty}^\infty \frac{d\theta_1}{2\pi}\dots \int_{-\infty}^\infty \frac{d\theta_n}{2\pi}
\left(\prod_{i=1}^n \frac{1}{1+e^{\eps_{a_i}(\theta_i)}}\right)
\varphi^+_{a_1a_2}(\theta_1+\theta_2)\varphi^-_{a_2a_3}(\theta_2-\theta_3)\dots 
\varphi^-_{a_na_1}(\theta_n-\theta_1)
\end{eqnarray}
The second summation runs over $a_i=1,2$, the pseudo-energies are given by
\begin{equation*}
  \eps_1(\theta)=\eps(\theta) \qquad \eps_2(\theta)=\eps(-\theta)
\end{equation*}
with $\eps(\theta)$ being the solution of \eqref{ujTBA} and the
kernels are defined as
\begin{equation*}
  \varphi^+_{jk}(\theta) = \left\{
    \begin{array}{ll}
      \varphi_{12}(\theta) \text{ for } j=k\\
  \varphi_{11}(\theta) \text{ for } j\ne k
    \end{array}
\right.
\qquad\text{and}\qquad
 \varphi^-_{jk}(\theta) = \left\{
    \begin{array}{ll}
      \varphi_{11}(\theta) \text{ for } j=k\\
  \varphi_{12}(\theta) \text{ for } j\ne k
    \end{array}
\right.
\end{equation*}
Equation \eqref{verygood} can be proven term by term using the
symmetry $\varphi_{jk}^\pm(\theta)=\varphi_{jk}^\pm(-\theta)$.

\subsection{The massless flow from tri-critical Ising to critical Ising}

The simplest non-trivial massless model is probably the scattering theory describing
the flow from the tri-critical Ising to the critical Ising model
\cite{Zamolodchikov:1991vx}. In this theory there is only one particle
species and the scattering is described by
\begin{equation}
  S_{LL}(\theta)=S_{RR}(\theta)=1,\qquad
S_{LR}(\theta)=-\tanh(\theta/2-i\pi/4)
\end{equation}
In \cite{Dorey:2009vg} $g$-function flows were studied between
different conformal boundary conditions of the UV and IR theories in
those cases where both the bulk and the boundary perturbations are
integrable and compatible with each other. There are two
such possibilities:
\begin{enumerate}
\item The flow from the boundary condition $(0+)$ of tri-critical
  Ising to $(+)$ in Ising
\item The flow from the boundary condition $(d)$ of tri-critical
  Ising to $(f)$ (free) in Ising
\end{enumerate}
Both flows are induced by the $\Phi_{13}$ perturbation on the boundary.
For the precise definition of
the boundary conditions we refer to \cite{Dorey:2009vg} and references therein.
Here we show that the results of
\cite{Dorey:2009vg} can be derived from our general formalism. Most
importantly, we present an all-orders proof of the
boundary-independent part of the $g$-function, which differs from the
massive version.

In the present case the integration kernels are given by
\begin{equation*}
  K_1(\theta,\theta')=-K_2(\theta,\theta')=\varphi(\theta+\theta')=\frac{1}{\cosh(\theta+\theta')}
\end{equation*}
The function $\sigma(\theta)$ reads
\begin{equation*}
  \sigma(\theta)=me^\theta+\Theta_{ab}(\theta)
\end{equation*}
where 
\begin{equation*}
  \Theta_{ab}(\theta)=-i\frac{d}{d\theta}\frac{1}{L}
 \log R_{ab}(\theta),\quad\quad R_{ab}(\theta)=R_a(\theta)R_b(\theta)S_{LR}(-2\theta) 
\end{equation*}
The possible reflection factors $R_{a}(\theta)$ and  $R_{b}(\theta)$ were specified in \cite{Dorey:2009vg}.

The TBA equation is given by
\begin{equation}
\label{ujTBA2}
    \eps(\theta)=\frac{1}{2} mR e^\theta -\int_{-\infty}^\infty
    \frac{d\theta'}{2\pi}\varphi(\theta+\theta')\log\big(1+e^{-\eps(\theta')}\big)
\end{equation}
Formula \eqref{Fmin} yields
\begin{equation*}
  F_{\rm min}=F_{\rm min}^p+F_{ab}
\end{equation*}
Here $F_{\rm min}^p$ is given by \eqref{masslessFminA} and $F_{ab}$ is
an $\ordo(1)$ piece given by
\begin{equation*}
  F_{ab}=-T
\int_{-\infty}^\infty \frac{d\theta}{2\pi} 
\Theta_{ab}(\theta)\log\big(1+e^{-\eps(\theta)}\big)
\end{equation*}
Putting everything together, equation \eqref{joZ} yields
\begin{equation}
  Z=\frac{\det\big(1+\hat P^+\big)}{\det\big(1-\hat P^+\big)}e^{-\beta
    F_{\rm min}^p-\beta F_{ab}}
\end{equation}
where the operator $\hat P^+$ acts on functions defined on $\valos$ as
\begin{equation}
\label{R}
  \big(\hat P^+
  (f)\big)(x)=\int_{-\infty}^\infty\frac{dy}{2\pi}
\varphi(x+y)   \frac{1}{1+e^{\eps(y)}} f(y)
\end{equation}
The ratio of Fredholm-determinants can be evaluated using
\eqref{logdet}:
\begin{equation*}
\begin{split}
&\log  \frac{\det\big(1+\hat P^+\big)}{\det\big(1-\hat P^+\big)}=
\sum_{n=1}^\infty \frac{1}{n}\Big(\text{Tr}  {(\hat P^+)}^n-\text{Tr} {(-\hat P^+)}^n\Big)=\\
&2\sum_{j=1}^\infty \frac{1}{2j-1} 
\int_{\valos^{2j-1}} 
\frac{d\theta_1}{2\pi}\dots \frac{d\theta_{2j-1}}{2\pi}
\left(\prod_{i=1}^{2j-1} \frac{1}{1+e^{\eps(\theta_i)}}\right)
\varphi(\theta_1+\theta_2)\varphi(\theta_2+\theta_3)\dots \varphi(\theta_{2j-1}+\theta_1)
\end{split}
\end{equation*}
We find the exact $g$-function 
\begin{equation*}
  \log g= \log g_{a}+\log g_0
\end{equation*}
where the boundary dependent part is 
\begin{eqnarray}
\label{kjl}
&&  \log g_a=\int_{-\infty}^\infty \frac{d\theta}{2\pi} 
\Big(\varphi_a(\theta)-\varphi(2\theta)\Big)
\log\big(1+e^{-\eps(\theta)}\big)
\end{eqnarray}
with
\begin{equation*}
\varphi_a(\theta)=  -i\frac{d}{d\theta} \log R_a(\theta)
\end{equation*}
and the boundary independent part is
\begin{eqnarray}
\label{masslessg1}
&&\log g_0=\\
&& \nonumber
\sum_{j=1}^\infty \frac{1}{2j-1} 
\int_{\valos^{2j-1}} 
\frac{d\theta_1}{2\pi}\dots \frac{d\theta_{2j-1}}{2\pi}
\left(\prod_{i=1}^{2j-1} \frac{1}{1+e^{\eps(\theta_i)}}\right)
\varphi(\theta_1+\theta_2)\varphi(\theta_2+\theta_3)\dots \varphi(\theta_{2j-1}+\theta_1)
\end{eqnarray}
Equation \eqref{masslessg1} is in agreement with the corresponding
formula of \cite{Dorey:2009vg}. However, \eqref{kjl} coincides with
the result of \cite{Dorey:2009vg} only in the case of
the flow to the $(f)$ boundary condition in the Ising model. In the
other case, namely the
flow from $(0+)$ in tri-critical Ising to $(+)$ 
in Ising there is a missing term $-\frac{1}{2}\log2$. 
This discrepancy
can be explained by the fact, that in the IR limit the boundary
condition corresponds to  a microscopic theory where the ground state
degeneracy of 2 is removed. The scattering theory describes the
variation of the $g$-function with respect to the 
temperature, therefore it is
natural to assume that the extra term has to be added not just in the
IR limit (which corresponds to zero temperature), but also for the whole RG flow. Therefore we write
\begin{equation*}
    \log g_a=-\frac{1}{2}\log2+ \int_{-\infty}^\infty \frac{d\theta}{2\pi} 
\Big(\varphi_a(\theta)-\varphi(2\theta)\Big)
\log\big(1+e^{-\eps(\theta)}\big)
\end{equation*}
which is in agreement with \cite{Dorey:2009vg}.

Finally we mention that  \eqref{masslessg1} follows from the general
formula \eqref{verygood} after substituting
\begin{equation*}
  \varphi_{11}(\theta)=0\qquad \varphi_{12}(\theta)=\varphi(\theta)
\end{equation*}
and making the appropriate change of variables.

\subsection{The massless flow $M_{3,5}+\Phi_{2,1}\to M_{2,5}$}

In \cite{Ravanini:1994pt} a simple massless scattering theory with one
particle species was proposed to describe the
flow from the minimal model $M_{3,5}$ to $M_{2,5}$ induced by the
perturbing field $\Phi_{2,1}$ \cite{martins-roaming}. In this model the scattering is
described by
\begin{equation}
\label{S35}
  S_{LL}(\theta)=S_{RR}(\theta)=S_{LY}(\theta),\qquad
S_{LR}(\theta)=S_{RL}(\theta)=\big(S_{LY}(\theta)\big)^{-1},
\end{equation}
where $S_{LY}(\theta)$ is the S-matrix of the massive Lee-Yang model
\cite{smirnov-reductions,Cardy_Mussardo__Lee_Yang} 
\begin{equation*}
  S_{LY}(\theta)=\frac{\sinh\theta+i\sin(\pi/3)}{\sinh\theta-i\sin(\pi/3)}
\end{equation*}
The massive Lee-Yang model is the $\Phi_{1,3}$ perturbation of the
minimal model $M_{2,5}$. In massless theories the LL and RR scattering
matrices are scale-invariant and they describe the IR limiting CFT;
this was the motivation for the choice of $S_{LL}$ and $S_{RR}$ in \eqref{S35}.

The possible reflection factors of this model have not yet been
written down. Nevertheless it is useful to derive the $g$-function,
leaving the factors $R_{a}(\theta)$ and $R_{b}(\theta)$ unspecified. The boundary
dependent part will be given by \eqref{masslessFab}; in the following
we concentrate on the boundary independent part.

Given the
scattering matrices \eqref{S35} the integral kernels are given by 
\begin{equation*}
\begin{split}
  K_1(\theta,\theta')=\varphi(\theta-\theta')-\varphi(\theta+\theta')\\
  K_2(\theta,\theta')=\varphi(\theta-\theta')+\varphi(\theta+\theta'),
\end{split}
\end{equation*}
where
\begin{equation*}
  \varphi(\theta)=\frac{d}{d\theta} \log S_{LY}(\theta)
\end{equation*}
The TBA equation reads
\begin{equation}
\label{ujTBA4}
    \eps(\theta)=\frac{1}{2} mR e^\theta -\int_{-\infty}^\infty
    \frac{d\theta'}{2\pi}
\big(\varphi(\theta-\theta')-\varphi(\theta+\theta') \big)
\log\big(1+e^{-\eps(\theta')}\big)
\end{equation}
One can use the general formula \eqref{verygood} to express the
boundary independent part of the $g$-function as 
\begin{equation*}
\begin{split}
&\log g_0=\frac{1}{2}\log  \frac{\det\big(1-\hat K_2\big)}{\det\big(1-\hat K_1\big)}=\\
 & -\frac{1}{2}
\sum_{n=1}^\infty \sum_{a_1\dots a_{n}} \frac{1}{n}
\int_{-\infty}^\infty \frac{d\theta_1}{2\pi}\dots \int_{-\infty}^\infty \frac{d\theta_n}{2\pi}
\left(\prod_{i=1}^{n} \frac{1}{1+e^{\eps_{a_i}(\theta_i)}}\right)
\varphi(\theta_1+\theta_2)\varphi(\theta_2-\theta_3)\dots
\varphi(\theta_n-\theta_1)
\end{split}
\end{equation*}
The summations run over $a_i=1,2$ and the pseudo-energies are given by 
\begin{equation*}
  \eps_1(\theta)=\eps(\theta) \qquad \eps_2(\theta)=\eps(-\theta)
\end{equation*}
where $\eps(\theta)$ is the solution of \eqref{ujTBA4}. Notice that
due to the specific form of the S-matrix \eqref{S35}
there is an overall factor of $(-1)$ as compared to the massive case
\eqref{vegso1}.

\section{Conclusions}

\label{concl}

We have studied the partition function of Bethe Ansatz
solvable models as a function of the volume and the temperature. 
We have shown how to obtain $O(L^0)$ pieces to the free energy in the
framework of the Thermodynamic Bethe Ansatz: our main result is
equation \eqref{joZ}. In addition to possible boundary dependent parts
incorporated in the ``minimal part'' of the free energy $F_{min}$ the
formula \eqref{joZ} involves two Fredholm-determinants which depend only on the scattering in the
bulk.  In relativistic boundary field theory these two pieces are
responsible for the
boundary-independent part of the $g$-function.
We have presented a new result
\eqref{mounteverest} which applies to massless relativistic theories
with arbitrary diagonal scattering in the bulk.
This formula could be used to study massless bulk-boundary flows along
the lines of \cite{Dorey:2009vg}.

Formula \eqref{joZ} can be applied in a very straightforward
way once the Bethe-Yang equations have been established. Therefore it
is very natural to conjecture that a similar result will hold in
theories with non-diagonal scattering. In these models the finite volume
quantization  proceeds through the
diagonalisation of the transfer matrix, which can be achieved by the
introduction of the so-called magnonic (or spin) particles
\cite{YangYang2,Zamolodchikov:1991vh,Zamolodchikov:1992zr,Ravanini:1992fi}. Once
this 
algebraic problem is solved, the derivation of the TBA follows
straightforwardly; a common property is that there are no energy-terms $e(\theta)$
associated to the magnonic modes. We believe that our arguments can be
applied to these Bethe Ansatz systems, particularly in those cases
when the TBA results in a finite set of equations.  The study of these
$g$-functions with magnonic modes is left for future work. 

Also, it would be interesting to study the flow of the excited states
quantities $G^\Psi(R)$ defined in \eqref{djvu}. In models with discrete symmetries some
of the excited states can be treated simply
by introducing real or complex fugacities $\lambda_a=e^{-\mu_a /T}$ for the
different particle types \cite{martins-exc-TBA,Fendley:1991xn}. We expect that
our present results will hold in these modified TBA systems. However,
the problem of particle-like excited states will probably require new
ideas. In \cite{Kormos:2010ae} it was shown that generally the amplitudes
$G^\Psi(R)$ include normalization factors which are analytic in $1/R$;
it is not yet clear whether these factors can be interpreted in the
framework of Thermodynamic Bethe Ansatz.

\vspace{0.7cm}

{\bf Acknowledgements:} 
We would like to thank L. Palla,
Z. Bajnok, M. Kormos and in particular R. Tateo and
G. Tak\'acs for encouraging and very helpful discussions. Also, we are grateful to
M. Kormos, G. Tak\'acs and G. Palacios for useful comments on the manuscript.

\vspace{2cm}

{\bf\Large Appendix}

\appendix

\section{Relations between the Fredholm-determinants}

Here we consider simple relations between the different
Fredholm-determinants introduced in the main text. 

The operators $\hat P^+$ and $\hat P^-$ act on functions defined on $\valos$ as
\begin{equation*}
  \big(\hat P^\pm
  (f)\big)(x)=\int_{-\infty}^\infty\frac{dy}{2\pi}\varphi(x\pm y)  \frac{1}{1+e^{\eps(y)}} f(y)
\end{equation*}
The function $\varphi(\theta)=\vartheta'(\theta)$ is the scattering
kernel. A very important property is that it is symmetric:
\begin{equation*}
  \varphi(\theta)=\varphi(-\theta)
\end{equation*}
The operators $\hat Q^+$ and $\hat Q^-$ act on
functions defined on $\valos^+$ as
\begin{equation*}
  \big(\hat Q^\pm
  (g)\big)(x)=\int_{0}^\infty\frac{dy}{2\pi}\big(\varphi(x-y)\pm \varphi(x+y)\big)
  \frac{1}{1+e^{\eps(y)}} g(y)
\end{equation*}
Let us decompose the real line as $\valos=\valos^++\valos^-$ (the
point $x=0$ has zero measure, therefore it is irrelevant in the
present context).
It is easy to see, that in this decomposition
the operators $\hat P^+$ and $\hat P^-$ can be written in
block form as
\begin{equation}
\label{jokis}
  \hat P^+=
  \begin{pmatrix}
    \hat B& \hat A\\
\hat A &   \hat B 
 \end{pmatrix}
\qquad
  \hat P^-=
  \begin{pmatrix}
     \hat A& \hat B\\
\hat B &   \hat A 
 \end{pmatrix},
\end{equation}
where $\hat A$ and $\hat B$ are integral operators on $\valos^+$ with
kernels $\varphi(x-y)$ and $\varphi(x+y)$, respectively.

The determinant of the operators $\hat 1-\hat P^{\pm}$ 
can be evaluated as
\begin{equation*}
  \det \big(  \hat 1-\hat P^+\big)=
\det \big(\hat 1 -(\hat A+\hat B)\big)
\det \big(\hat 1 +(\hat A-\hat B)\big)
\end{equation*}
\begin{equation*}
  \det \big(  \hat 1-\hat P^-\big)=
\det \big(\hat 1 -(\hat A+\hat B)\big)
\det \big(\hat 1 -(\hat A-\hat B)\big)
\end{equation*}
Using the relations
\begin{equation*}
\hat A=\frac{\hat  Q^++\hat  Q^-}{2}\qquad
\hat B=\frac{\hat  Q^+-\hat  Q^-}{2}
\end{equation*}
one gets
\begin{equation}
\label{faktorizalodik2}
  \det \big(  \hat 1-\hat P^+\big)=
\det \big(  \hat 1-\hat Q^+\big) \det \big(  \hat 1+\hat Q^-\big)
\end{equation}
\begin{equation}
\label{faktorizalodik}
  \det \big(  \hat 1-\hat P^-\big)=
\det \big(  \hat 1-\hat Q^+\big) \det \big(  \hat 1-\hat Q^-\big)
\end{equation}
The relation \eqref{faktorizalodik} was used in the main text to prove the
equivalence of the expressions \eqref{g-vegso1} and
\eqref{g-vegso0}. We wish to mention that eqs. \eqref{faktorizalodik2}-\eqref{faktorizalodik} can
be proven alternatively term by term using formula \eqref{logdet}.

\addcontentsline{toc}{section}{References}
\bibliography{pozsi-general}
\bibliographystyle{utphys}

\end{document}